\documentclass[twocolumn,english,prx,superscriptaddress]{revtex4-1}
\usepackage{color}
\usepackage[utf8]{inputenc}
\usepackage{graphicx}
\usepackage[T1]{fontenc}
\usepackage{float}
\usepackage{natbib}
\usepackage{textcomp}
\usepackage{amsmath}
\usepackage{amssymb}
\usepackage{graphicx}
\usepackage{esint}
\usepackage{natbib}
\usepackage{xcolor}
\usepackage{multirow}
\usepackage{ulem}

\begin{document}
\title{Effective Land\'{e} factors for an electrostatically defined quantum point contact in silicene}

\author{Bart\l{}omiej Rzeszotarski}

\affiliation{AGH University of Science and Technology, Faculty of Physics and
Applied Computer Science,\\
 al. Mickiewicza 30, 30-059 Kraków, Poland}

\author{Alina Mre\'n{}ca-Kolasi\'n{}ska}

\affiliation{AGH University of Science and Technology, Faculty of Physics and
Applied Computer Science,\\
 al. Mickiewicza 30, 30-059 Kraków, Poland}
 
\affiliation{Department of Physics, National Cheng Kung University, Tainan 70101, Taiwan}

 \author{Fran\c{c}ois M. Peeters} 
 
\affiliation{Universiteit Antwerpen, Departement Fysica, Groenenborgerlaan 171, B-2020 Antwerpen, Belgium}

\author{Bart\l{}omiej Szafran}

\affiliation{AGH University of Science and Technology, Faculty of Physics and
Applied Computer Science,\\
 al. Mickiewicza 30, 30-059 Kraków, Poland}

\begin{abstract}
The transconductance and effective Land\'{e} $g^*$ factors 
for a quantum point contact defined in silicene by the electric field of a split gate is investigated. 
The strong spin-orbit coupling in buckled silicene reduces the 
 $g^*$ factor for in-plane magnetic field from the nominal value 2 to around 1.2 for the first- to 0.45 for the third conduction subband. However, for perpendicular magnetic field we observe an enhancement of $g^*$ factors for the
first subband to 5.8 in nanoribbon with zigzag and to 2.5 with armchair edge. The main contribution to the Zeeman splitting comes from the intrinsic spin-orbit coupling defined by the Kane-Mele form of interaction.
\end{abstract}
\maketitle
\section{Introduction}

Quantum point contacts (QPC) in spin-orbit-coupled semiconductors are elementary elements in the construction of spin-active devices due to their ability to enchance the effective Lande factor $g^*$ \cite{Thomas96}. In the absence of external magnetic field QPC system with strong spin-orbit interaction can work as a spin filter \cite{Bhandari2013,Nowak2013,Kim2012,Aharony2008,Eto2005}. Spin orbit interactions  due to the crystal lattice asymmetry and external electric fields introduce effective magnetic fields \cite{Meier2007, km, cum} for the flowing electrons. The orientation of an external magnetic field (in-plane or out-of-plane) has a strong impact on conductance due to the spin spatial anisotropy of the spin-orbit field \cite{Goulko2014,Pershin2004,Scheid2008} whichd has been observed experimentally \cite{Martin10,Lu2013} 
by splitting the transconductance lines.
In systems with strong spin orbit interaction the anisotropy is very strong, e.g. in InSb QPCs \cite{InSbqpc} the in-plane $|g^*|=26$ and out-of-plane is two times higher, $|g^*|=52$ for the lowest conducting subband.
On the other hand in materials with low intrinsic spin-orbit coupling such as pristine graphene, the $g^*$ value is  $\simeq 2$ as for free electrons \cite{graphene_g1,graphene_g2,graphene_g3}. 
In bilayer graphene (BLG) structures quantum point contacts can be formed electrostatically \cite{Overweg2017,Overweg18,Banszerus20,BLG_qpc20,Kraft18} due to the opening of a band gap that can be tuned by a perpendicular electric field \cite{Zhang2009,Oostinga2007,Castro2007,Ohta2006,bigfz}. The spin $g^*$ is still $\simeq 2$ in bilayer graphene QPC \cite{BLG_qpc20}, however the valley $g$ factor can be tuned and used as an additional degree of freedom. In the silicene \cite{Liu11, Liu15, chow, Tao15}, a graphene-like honeycomb structure, two sublattices displaced in $z$ direction introduce strong intrinsic spin orbit interaction \cite{Ezawa}. Additionaly, the band gap in silicene can be electrostatically modified by external gates \cite{Drummond12, Tsai13, shak15,ni} that makes it a good candidate for a spin-active device.

In this paper we present a numerical calculation of the effective Land\'{e} $g^*$ factors for silicene using the transconductance lines according to a standard experimental procedure of determining the $g^*$ values \cite{Danneau2006,Martin2008,Martin10,Lu2013,Kolasinski16}. We test the $g^*$ anisotropy by dependence on the orientation of the external magnetic field. We discuss impact of the SO interaction on $g^*$ value for in-plane and out-of-plane magnetic field.

\section{Theory}

\begin{figure}[htbp]
\centering
\includegraphics[width=0.45\textwidth, clip, trim=2cm 2cm 2cm 4cm]{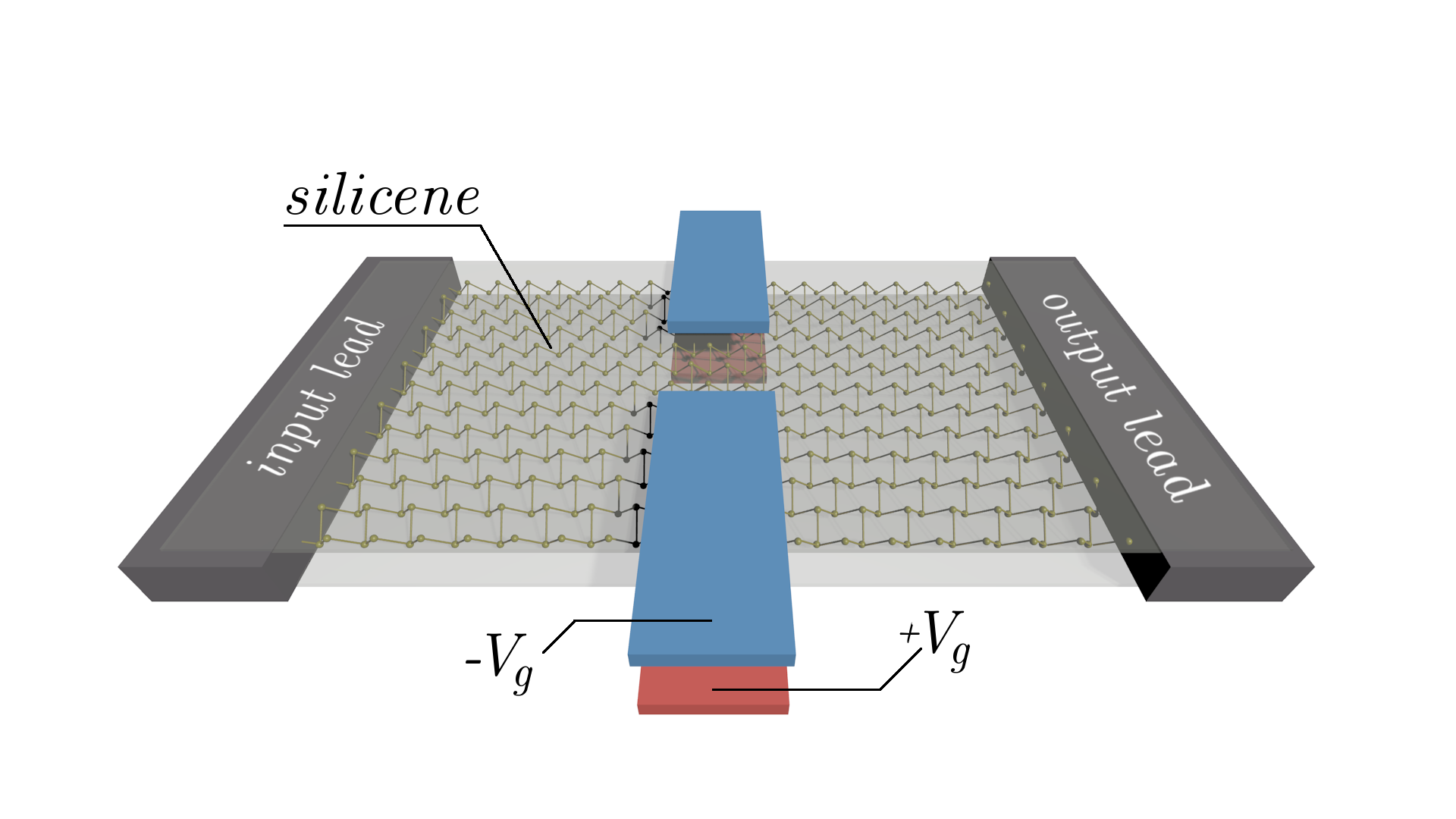}
\caption{Schematic view of the system. A silicene nanoribbon is sandwiched between dielectric layers. Two pairs of external gates provides electric potential $V_g$ equalizing the effective energy on both sublattices}
\label{fig:sch}
\end{figure}

\begin{figure}[htbp]
\centering
\includegraphics[width=0.45\textwidth, clip, trim=0cm 0cm 0cm 0cm]{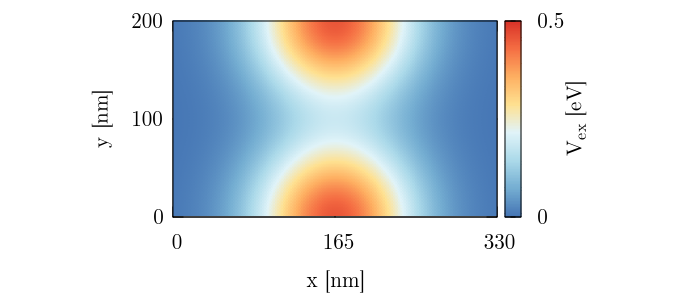}
\includegraphics[width=0.40\textwidth, clip, trim=0cm 0cm 0cm 0cm]{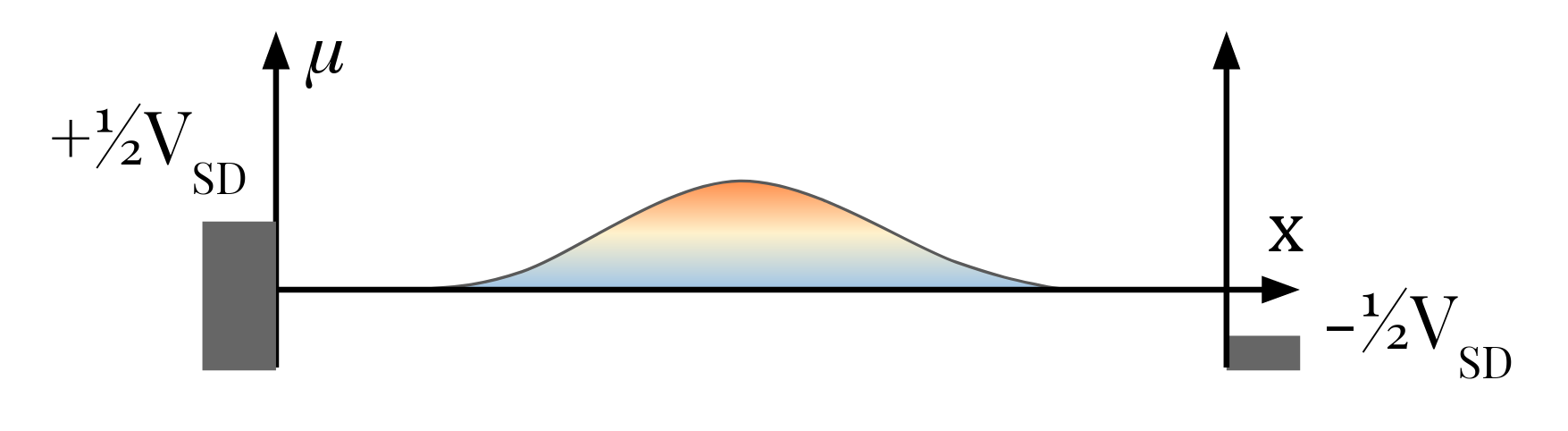}
\caption{ (a) Map of the external electric potential for $V_g=0.5$ eV, in the system of size 200 nm by 330 nm. Two Gaussian 2D potentials are used to form the quantum point contact in the middle of the silicene nanoribbon. (b) Profile of the QPC. The occupied states below chemical potential ($\pm V_{SD}/2$) are marked on both sides by rectangles with the assumption of a symmetric drop along the device. }
\label{fig:vexmap}
\end{figure}

We consider a device with a quantum point contact defined in silicene nanoribbon [Fig.~\ref{fig:sch}]. The QPC profile is defined by external split gates at voltages $\pm V_g/(-e)$ which induce a potential $V_{ex}$ at both sublattices equally forming QPC profile. In our calculations we assume a model potential profile given by a Gaussian \citep{Petrovic15} 
\begin{equation}
V_{Gauss}(x,y;x_{0},y_{0},\Delta x,\Delta y) = V_g e^{\frac{-(x-x_0)^2}{(2\Delta x)^2}}e^{\frac{-(y-y_0)^2}{(2\Delta y)^2}}
\end{equation}
and model the QPC with
\begin{align}
V_{ex} = & {\ }V_{Gauss}(x,y;165\text{ nm},0\text{ nm},40\text{ nm},40\text{ nm}) \nonumber \\
& + V_{Gauss}(x,y;165\text{ nm},200\text{ nm},40\text{ nm},40\text{ nm}).
\end{align}
The higher the applied gate voltage, the narrower the conductive channel in the center of the QPC. Figure \ref{fig:vexmap} presents example of gate energy distribution within silicene for the specific case where $V_g=0.5$ eV [Fig.~\ref{fig:vexmap}(a)], and the profile of the potential [Fig.~\ref{fig:vexmap}(b)]. 

\subsection{Hamiltonian}

We use the tight-binding Hamiltonian \cite{Liu}:
\begin{align}
H_{eff}=&-t\sum_{\langle k,j\rangle, \chi }  c_{k \chi}^\dagger c_{j \chi}+e \mathbb{E}_z \sum_{k,\chi} \gamma_k c^\dagger_{k,\chi}c_{k,\chi},  \nonumber \\ 
& + \frac{1}{2}g\mu_B\sum_{k,\chi,\varrho}c^\dagger_{k,\chi}c_{k,\varrho}(\mathbf{B}\cdot\pmb{\sigma})_{\chi\varrho} + H_{SO},  
\label{eq:heff}
\end{align}
where we use creation ($c_{k \chi}^\dagger$) and annihilation ($c_{k \chi}$) operators for an electron on site $k$ with spin $\chi$. Ions in the nearest neighborhood are specified by $\langle k,j\rangle$. $t=1.6$ eV is the hopping parameter  \cite{Liu,Ezawa} and $e$ is the elementary electric charge. The $\mathbb{E}_z$ term describes external perpendicular electric field with a factor $\gamma_k=\frac{1}{2}\cdot 0.46$ \AA{} that determines the offset in the sublattices. The penultimate term introduces the external magnetic field $\mathbf{B}=[b_x,b_y,b_z]$ to the system, where $\pmb{\sigma}=[\sigma_x,\sigma_y,\sigma_z]$ is a vector of Pauli matrices. We use the Landé factor $g=2$ for electrons in silicene along with Bohr magneton constant $\mu_B$.
The last term describes the spin-orbit part of the effective Hamiltonian $H_{eff}$:
\begin{align}
H_{SO} = &{\ }t_{KM} \sum_{\langle \langle k,j\rangle \rangle \chi, \varrho } \nu_{kj} c^\dagger_{k\chi} \sigma^{z}_{\chi,\varrho}c_{j\varrho} \nonumber \\
&+t_{R} \sum_{\langle \langle  k,j \rangle \rangle \chi,\varrho } \mu_{kj} c^\dagger _{k\chi}\left(\pmb{\sigma}\times\pmb{d}_{kj} \right)^z_{\chi\varrho} c_{j\varrho}, 
\label{eq:h0}
\end{align}
where the first part describes the intrinsic spin-orbit coupling in Kane-Mele (KM) form \cite{km,km2} with  $t_{KM}=i\frac{\lambda_{SO}}{3\sqrt{3}}$ and $\lambda_{SO}=3.9$ meV, while the second term is an intrinsic Rashba spin-orbit interaction $t_{R}=-i\frac{2}{3}\lambda_{R}^{}$ with $\lambda_{R}=0.7$ meV \cite{Liu,Ezawa}. The summation in both cases runs over next-nearest neighbor ions $\langle \langle k,j\rangle \rangle$, where $\mu_{kj}$ is +1 or -1 for sublattice $A$ and $B$, respectively. The $\nu_{kj}=+1$  ($-1$) for the counterclockwise (clockwise) hopping from $j$ to $k$ ion, where $\pmb{d}_{kj}$ is a vector pointing from ion $k$ to ion $j$. The lattice constant $a=3.86$ \AA{}.
To calculate the total conductance we use the Landauer formalism

\begin{equation}
G = \frac{e^2}{\hbar} T_{total}(E_F) =  \frac{e^2}{\hbar} \sum^N_m T_m,
\label{eq:glan}
\end{equation}
where $N$ is the total number of propagating modes and $T_m$ is the transmission probability of the $m$th mode from the input to the output lead. 
We use quantum transmitting boundary method to solve the scattering problem \cite{Kolasinski16}.
For finite potentials of source and drain the current is calculated as follows:

\begin{equation}
I(V_{SD}; T= 0 ) = \frac{e}{h} \int^{+e\frac{V_{SD}}{2}}_{-e\frac{V_{SD}}{2}} T_{total}(E_F + E)dE,
\label{eq:curr_conv}
\end{equation}
with the assumption of a drop of the potential along the device for nonequivalent chemical potential of the leads [Fig. \ref{fig:vexmap}(b)]. 
With a nonzero bias, we use the formula for the conductance

\begin{equation}
G = \frac{d I (V_{SD})}{d V_{SD}}
\label{eq:g_conv}
\end{equation}
and we define the transconductance $dG/dV_g$ as a second mixed derivative of the current,

\begin{equation}
\frac{dG}{dV_g} = \frac{d^2I(V_{SD})}{dV_{SD}dV_g}.
\label{eq:trans_conv}
\end{equation}

The classical procedure of calculating the effective Land\'{e} g* factors from transconductance is based on compensation of the Zeeman splitting by application of source-drain bias into the system along with energy modulation from the gate potential \cite{Danneau2006,Martin2008,Martin10,Lu2013}. The gate-to-energy conversion factor can be determined for each subband from the slope of the transconductance lines in $B=0$ according to the formula:

\begin{equation}
\xi_{m} = \frac{1}{2} \frac{dV_{SD}}{dV_g}
\label{eq:cf}
\end{equation}
where the $1/2$ factor results of source-drain potential shift that is equal to half of the applied bias: 1/2 $V_{SD}$.
The final step in the procedure of finding $g^*$ factors is to evaluate the transconductance as a function of magnetic field $B$, and for each subband to find the susceptibility as the derivative $\frac{d(\Delta V_g(B))}{dB}$. Then the effective Land\'{e} factor for $m$th subband is given by

\begin{equation}
g_m^* = \frac{1}{\mu_B} \frac{d(\Delta V_g(B))}{dB} \xi_{m}.
\label{eq:gstar}
\end{equation}

\section{Results}
To reduce the numerical cost of the calculations we use the scaling method \cite{Liu15} with a scaling factor $s_f=4$, that gives new crystal lattice constant $a_s=a\cdot s_f$ along with new hopping parameter $t_s= \frac{t}{s_f}$. We replace $a$ to $a_s$ and $t$ to $t_s$ in Hamiltonian (\ref{eq:h0}).  All the results below are presented for the Fermi energy $E_F = 0.07$ eV, if not stated otherwise. 

\subsection{Band structure}
\begin{figure}[htbp]
\centering

\includegraphics[width=0.24\textwidth, clip, trim=0cm 0cm 0cm 0cm]{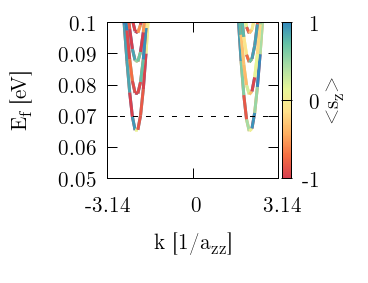}\put(-84,40){a)}\includegraphics[width=0.24\textwidth, clip, trim=0cm 0cm 0cm 0cm]{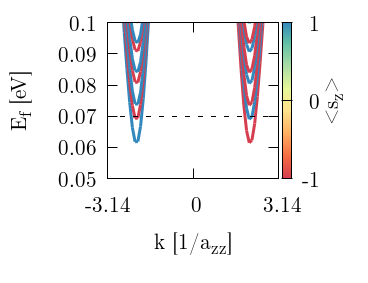}\put(-84,40){b)}\\
\includegraphics[width=0.24\textwidth, clip, trim=0cm 0cm 0cm 0cm]{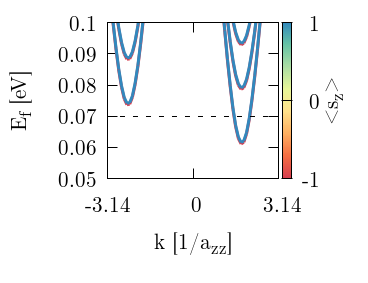}\put(-84,40){c)}\includegraphics[width=0.24\textwidth, clip, trim=0cm 0cm 0cm 0cm]{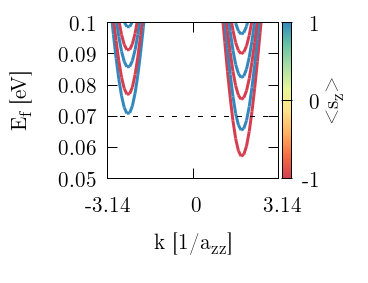}\put(-84,40){d)}
\caption{ Band structure of silicene nanoribbon with zigzag edges calculated for the center of the QPC constriction with $V_g=0.15$ eV, with spin-orbit interactions neglected (a,c) and included (b,d). Magnetic field is equal to $B_z=0$ (a,b) and $B_z=2$ T (c,d). Black dashed line denotes the Fermi energy $E_F=0.07$ eV. The color bar indicates the mean value of the spin projection along $z$ axis. The $a_{zz}=3a_s$ is the zigzag lattice constant. }
\label{fig:zigbands}
\end{figure}

\begin{figure}[htbp]
\centering

\includegraphics[width=0.24\textwidth, clip, trim=0cm 0cm 0cm 0cm]{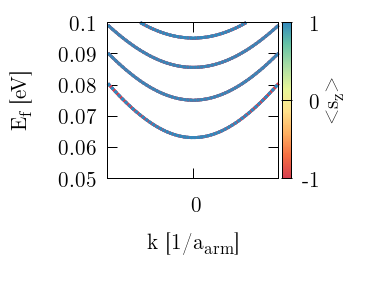}\put(-84,40){a)}\includegraphics[width=0.24\textwidth, clip, trim=0cm 0cm 0cm 0cm]{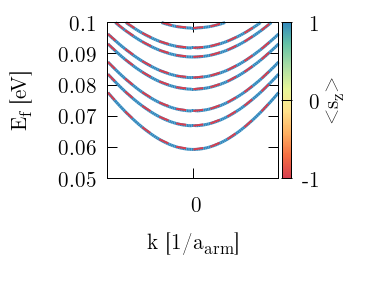}\put(-84,40){b)}\\
\includegraphics[width=0.24\textwidth, clip, trim=0cm 0cm 0cm 0cm]{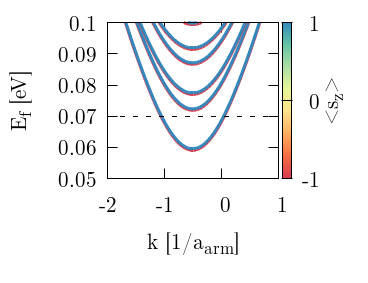}\put(-84,40){c)}\includegraphics[width=0.24\textwidth, clip, trim=0cm 0cm 0cm 0cm]{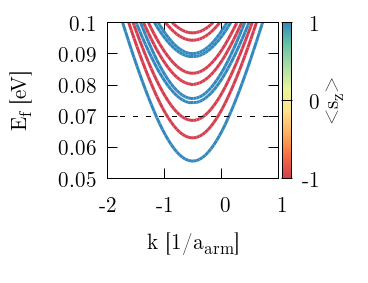}\put(-84,40){d)}
\caption{ Same as Fig.~\ref{fig:zigbands} but for a nanoribbon with armchair edges. Armchair lattice period is equal to $a_{arm}=\frac{6a_s}{\sqrt{3}}$.}
\label{fig:armbands}
\end{figure}

For the constriction center of the QPC we calculate the band structure for two different edge types: armchair and zigzag. In the zigzag case when spin-orbit interactions are omitted ($H_{SO}=0$) we observe spin-degenerate subbands at $\mathbf{B}=0$ [Fig.~\ref{fig:zigbands}(a)] for both valleys $K'$ and $K$, while this degeneracy is lifted upon applying an external magnetic field perpendicular to the sample [Fig.~\ref{fig:zigbands}(c) for $B_z=2$ T] that slightly splits the spin-states and shifts the subbands higher for $K$ and lower for $K'$.
When all spin-orbital interactions are included ($H_{SO}\neq 0$) then degeneracy is lifted even at $\mathbf{B}=0$, since the Zeeman-like SO interaction in KM interaction \citep{km,km2} introduces an effective magnetic field with an amplitude along the $z$ axis that splits the spin-states in the subbands [Fig.~\ref{fig:zigbands}(b)]. Up ($\uparrow$) spin states decrease their energy in the $K$ valley and increase in the $K'$ valley, while down ($\downarrow$) spin states shift in an opposite way. Applying an external perpendicular magnetic field in the case with SO interactions taken into account changes the energy gap in the same manner as with SO interactions omitted [Fig.~\ref{fig:zigbands}(d)]. We observe an analogous behavior for the armchair type of edges [Fig.~\ref{fig:armbands}]. 

\begin{figure}[htbp]
\centering
\includegraphics[width=0.4\textwidth, clip, trim=0cm 0cm 0cm 0cm]{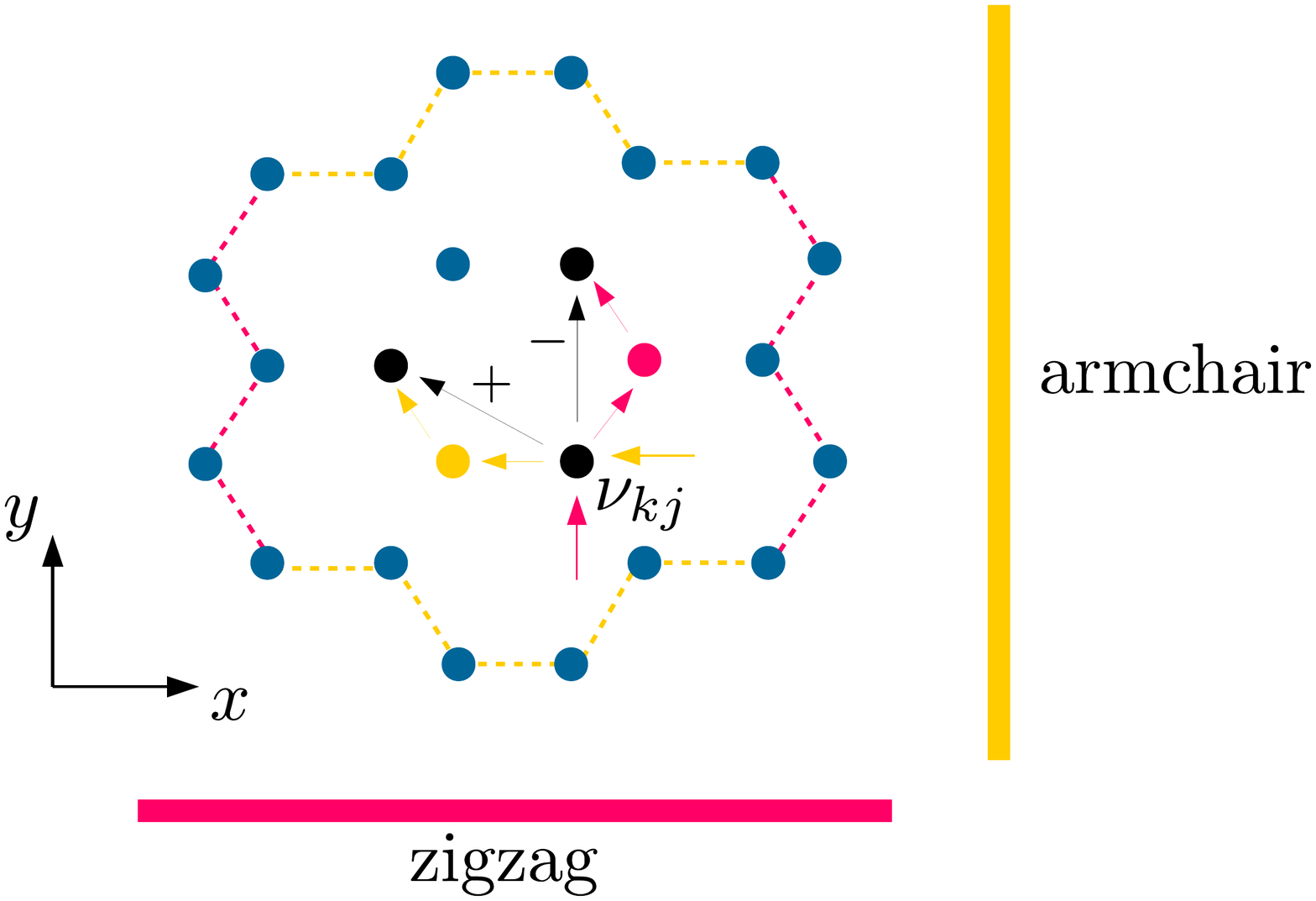}
\caption{ Schematic view of the $\nu_{kj}$ (Eq. \ref{eq:h0}) sign for two different edge types of silicene. For the same considered atom $j$ paths to the next-nearest neighbor $k$ differ in zigzag and armchair configuration and produce opposite sign of the local effective magnetic field resulting from KM term. This will produce mirrored image of spin signs in subband structures for armchair and zigzag.}
\label{fig:kmflow}
\end{figure}

\subsection{Conversion factors}

\begin{figure}[htbp]
\centering
\includegraphics[scale=0.18, clip, trim=0cm 2.7cm 7.1cm 0cm]{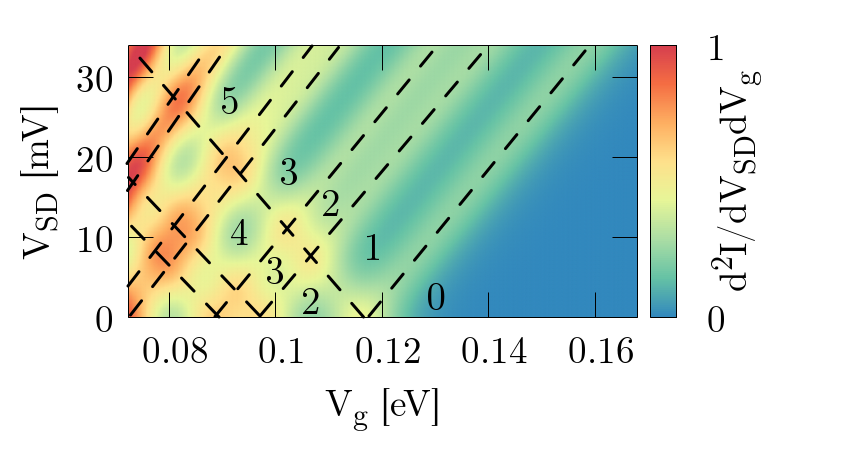}\put(-70,68){armchair}\includegraphics[scale=0.18, clip, trim=2.5cm 2.7cm 0cm 0cm]{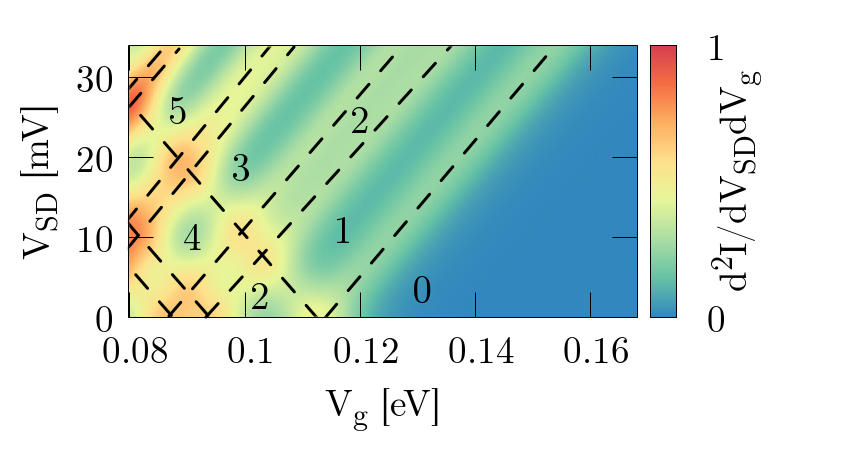}\put(-70,68){zigzag}\\
\includegraphics[scale=0.18, clip, trim=0cm 0cm 7.1cm 1cm]{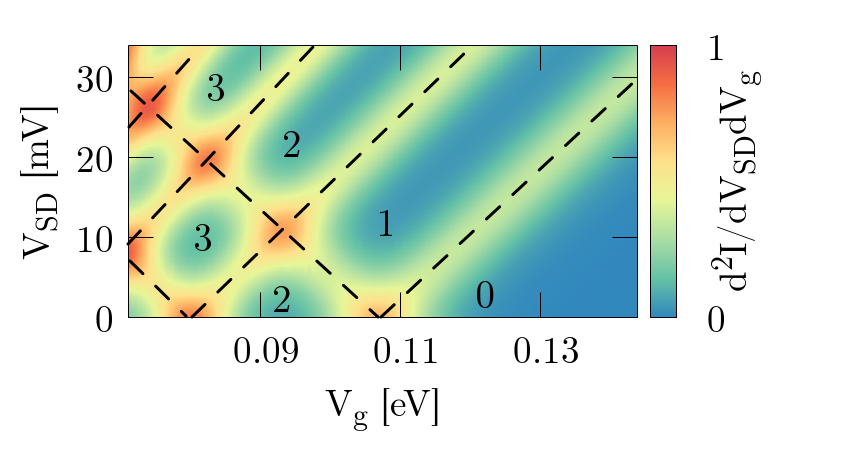}\includegraphics[scale=0.18, clip, trim=2.5cm 0cm 0cm 1cm]{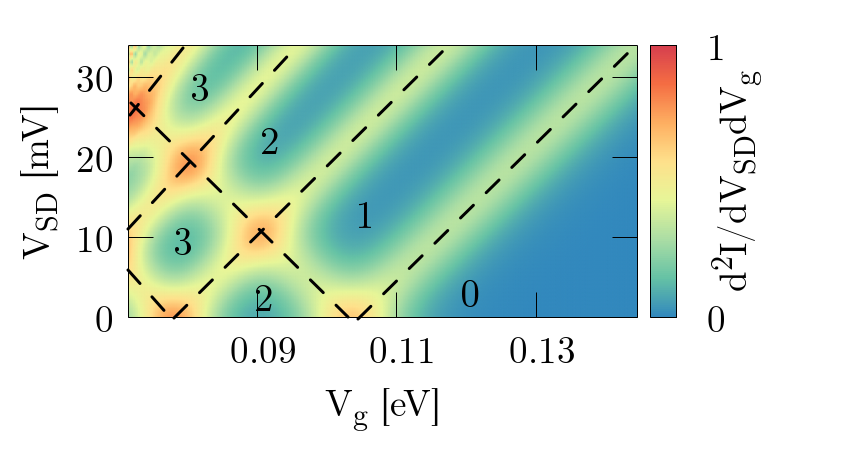}

\caption{ The transconductance $d^2I/dV_{SD}dV_{g} = \frac{dG}{dV_g}$ for armchair (left column) and zigzag (right column) nanoribbons with SO interactions (top row) and without (SO interactions neglected). Dashed lines denote the conductance levels (marked by numbers) and are used to calculate gate-voltage to energy conversion factors. }
\label{fig:efmap}
\end{figure}

We calculate the transconductance with a bias $V_{SD}$ applied using Eqs.~(\ref{eq:curr_conv}-\ref{eq:trans_conv}). Fig.~\ref{fig:efmap} presents maps of transconductance for armchair and zigzag edges for the cases -- with spin-orbit interactions included or neglected in the Hamiltonian. The dependence of the compensation of the Zeeman splitting by source-drain bias $V_{SD}$ on the gate voltage $V_G$ are marked by straight dashed lines for each subband with $H_{SO}=0$ [Fig.~\ref{fig:efmap}(c,d)]. For each subband we calculate the conversion factors (Eq.~\ref{eq:cf}) from the slope of the corresponding line $\frac{dV_{SD}}{dV_g}$ for both types of nanoribbons: armchair and zigzag. Results are presented in Tab.~\ref{tab:conv}. 

In the case with spin-orbit interactions taken into consideration [Fig.~\ref{fig:efmap}(a,b)] we observe twice more subbands that emerge from splitting caused by the Zeeman-like part of the intrinsic SO coupling.

\begin{table}[h]
\caption{Conversion factors for the first three subbands calculated from the transconductance (Fig.~\ref{fig:efmap}) without spin-orbit interaction.}
\label{tab:conv}
\begin{tabular}{|c|c|c|c|}
\hline
 & $\xi_{1}$ & $\xi_{2}$ & $\xi_{3}$ \\ \hline
armchair & 0.40 & 0.42 & 0.47 \\ \hline
zigzag & 0.41 & 0.44 & 0.49  \\ \hline
\end{tabular}
\end{table}

\subsection{Effective Land\'{e} factors}

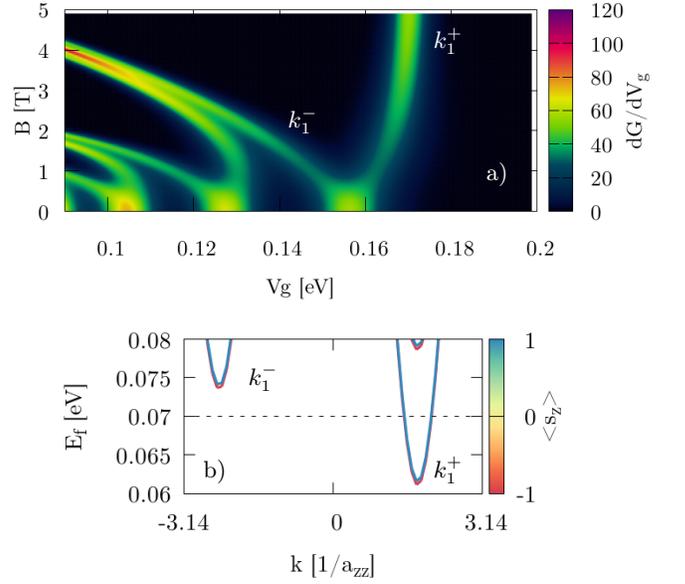
\begin{figure}[htbp]
\centering

\includegraphics[width=0.5\textwidth, clip, trim=0cm 0cm 0cm 1cm]{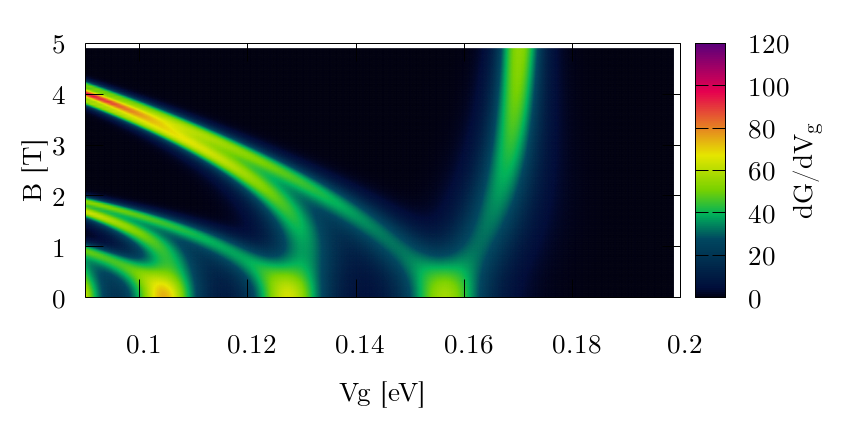}\put(-145,70){\color{white}$k^-_1$}\put(-90,100){\color{white}$k^+_1$}\put(-70,50){\color{white}a)}\\
\includegraphics[width=0.40\textwidth, clip, trim=0cm 0cm 0cm 0cm]{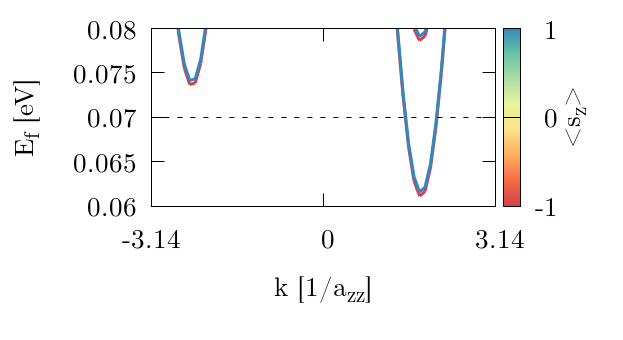}\put(-130,85){$k^-_1$}\put(-60,50){$k^+_1$}\put(-147,50){b)}\\
\caption{ (a) Transconductance for the  perpendicular $B\bot$ orientation of the magnetic field with SO interactions neglected ($H_{SO}=0$). 
Smooth transconductance peaks correspond to new conductive states in the subbands (two first marked as $k^+_1$ and $k^-_1$). Each subband includes two opposite spin states which are close enough not to be seen as separate peak [see zoom in plot (b)]. Results are for zigzag edges. }
\label{fig:zignoso}
\end{figure}

We calculate the transconductance by taking the derivative of the $G$ maps with respect to $V_g$.
First we consider perpendicular magnetic field $B_\bot = \mathbf{B} = (0,0,B_z)$. We present results for $H_{SO}=0$ [Fig.~\ref{fig:zignoso}] only for the zigzag nanoribbon since for the armchair system similar results are obtained. The two separate spin-states cannot be distinguished from the transconductance map so the calculation of the effective Land\'{e} factor is not possible in the standard way. However, we are able to identify valley and spin-state from the band structure in Fig.~\ref{fig:zignoso}(b). Upon subtraction of the energies at different $B_z$/$V_g$ values we find $g^*=2.0$, which agrees with the expected value $g=2$ for electrons in silicene.

\begin{figure}[htbp]
\centering

\includegraphics[width=0.5\textwidth, clip, trim=0cm 0cm 0cm 1cm]{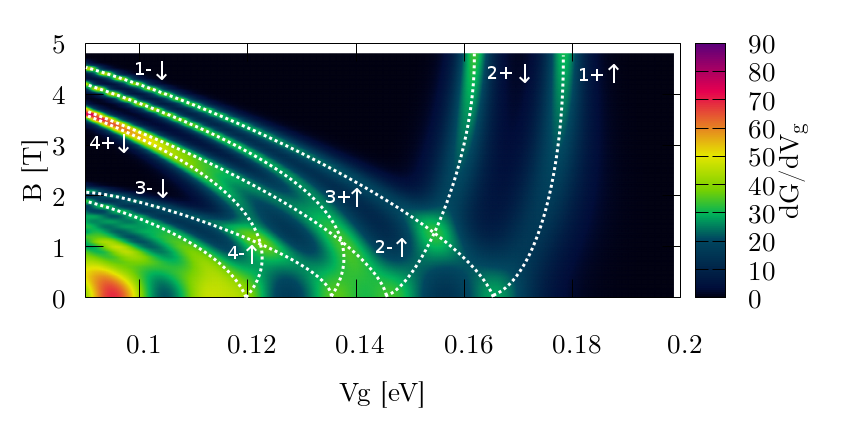}\put(-95,40){\color{white}armchair}\put(-80,70){\color{white}$B\bot$}\\
\includegraphics[width=0.5\textwidth, clip, trim=0cm 0cm 0cm 1cm]{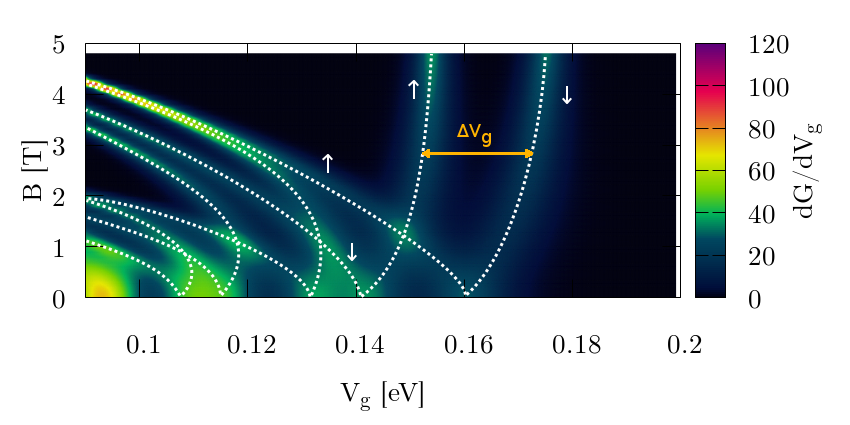}\put(-95,40){\color{white}zigzag}\put(-80,70){\color{white}$B\bot$}
\caption{ Transconductance for perpendicular $\mathbf{B}=[0,0,B_z]$ orientation of external magnetic field applied to the system with active SO (nonzero terms in Hamiltonian). Numbers denote the subband index in $k>0$ valley (+) and $k<0$ (-) with spin $\uparrow$ ($\downarrow$) marked by white dotted lines. The $\Delta V_g$ term is calculated between two spin-correlated subbands within the same valley band. }
\label{fig:outplane_cond}
\end{figure}

For nonzero $H_{SO}$ in the Hamiltonian (\ref{eq:heff}) the intrinsic SO interaction in the Zeeman-like form separates the spin states and now they can be easily distinguished in the transconductance map [Fig.~\ref{fig:outplane_cond}] when external magnetic field $B_z$ is applied. Identification of the subband and valley number comes from the band structure of the zigzag nanoribbon [Fig.~\ref{fig:zigbands}]. Calculating the slope of $d(\Delta V_g(B))/dB$ (marked by dashed lines) and using Eq.~(\ref{eq:gstar}) with conversion factors [Tab.~\ref{tab:conv}] we obtain $g^*_1=5.8$, $g^*_2=13.3$ for the case with zigzag edges and  $g^*_1=2.5$, $g^*_2=14.0$ for armchair edges [Tab.~\ref{tab:gstar}]. The difference comes directly from the geometry (Fig. \ref{fig:kmflow}) where $\nu_{kj}$ in KM term defines the sign of an additional energy to spin states. 
 Applying external magnetic field compensates this energy if its direction agrees with the emerged local magnetic field $\left[ \frac{d(\Delta V_g(B))}{dB} > 0\ \text{for}\ K'\ (k>0)\ \text{in zigzag}\ \right]$, or forfeit for the opposite directions $\left[ \frac{d(\Delta V_g(B))}{dB} < 0\ \text{for}\ K\ (k<0)\ \text{in zigzag}\ \right]$. We obtain mirrored behavior in armchair nanoribbons due to the $\nu_{kj}$ sign. In Fig. \ref{fig:fity}(c,d) we see that the slope for the first subband ($N_1$) is positive in the zigzag structure and negative in the armchair. Small difference in $g^*_2$ (slopes $N_2$ in Fig. \ref{fig:fity}(c,d)) for armchair and zigzag comes from the fact that deeper conductive bands have higher energy and additional fraction that comes from KM term is less significant in this scenario.

\begin{figure}[htbp]
\centering

\includegraphics[width=0.5\textwidth, clip, trim=0cm 0cm 0cm 0cm]{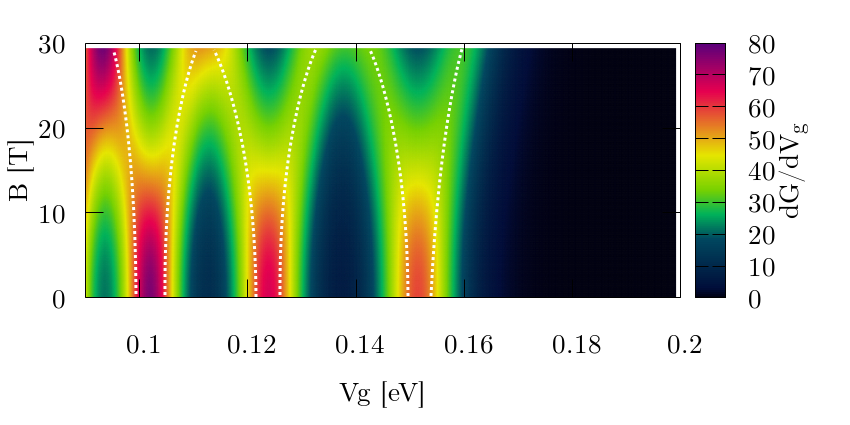}\put(-65,45){\color{white}a)}\put(-220,120){$H_{SO}=0$}\\
\includegraphics[width=0.5\textwidth, clip, trim=0cm 0cm 0cm 0cm]{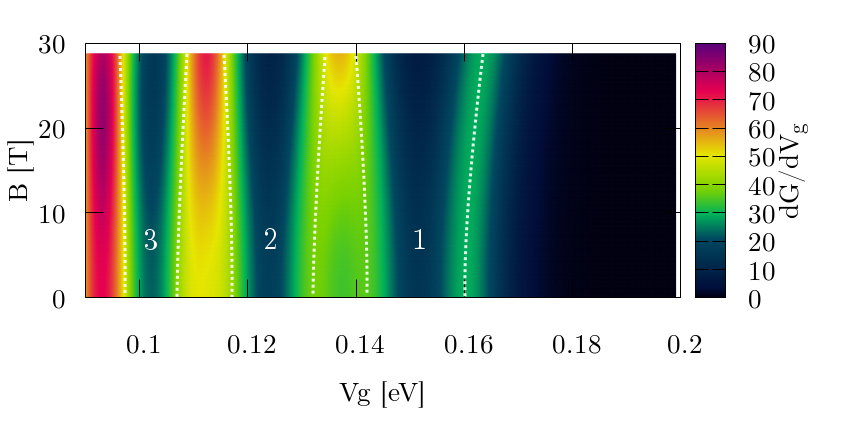}\put(-65,45){\color{white}b)}\put(-220,120){$H_{SO}\neq 0$}\\
\includegraphics[width=0.25\textwidth, clip, trim=0cm 0cm 0cm 0cm]{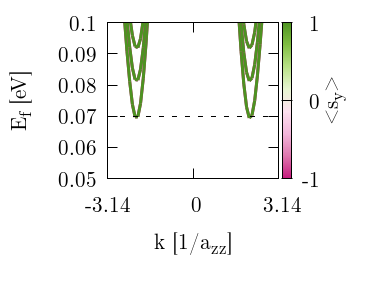}\includegraphics[width=0.25\textwidth, clip, trim=0cm 0cm 0cm 0cm]{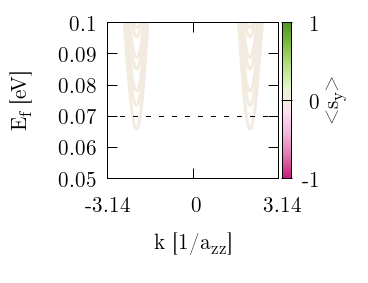}\put(-215,40){c)}\put(-87,40){d)}\put(-200,40){$B_y$=0}\put(-72,40){$B_y$=0}\\
\includegraphics[width=0.25\textwidth, clip, trim=0cm 0cm 0cm 0cm]{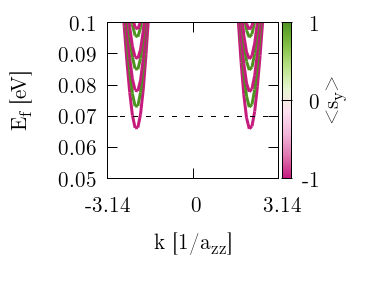}\includegraphics[width=0.25\textwidth, clip, trim=0cm 0cm 0cm 0cm]{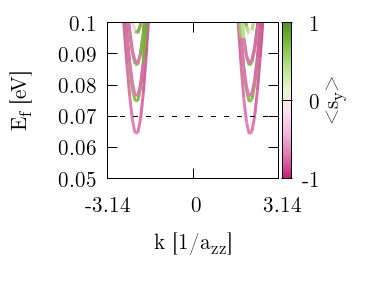}\put(-215,40){e)}\put(-87,40){f)}\put(-203,40){$B_y$=30 T}\put(-78,40){$B_y$=30 T}
\caption{ (a,b) Transconductance for the in-plane $\mathbf{B}=[0,B_y,0]$  orientation of the magnetic field for zigzag edges. Double-spin states (of the same sign from both valleys) at $E_F$  that enhance the conductance are marked by white dashed lines along the peaks. (c-f) Band structure for the center of the constriction. Colorbar indicates expected value of the spin projected to the $y$ axis. Left column (a,c,e) corresponds to calculations with neglected SO part of the Hamiltonian, while right column (b,d,f) are with included SO part, respectively.   }
\label{fig:inplane_cond}
\end{figure}

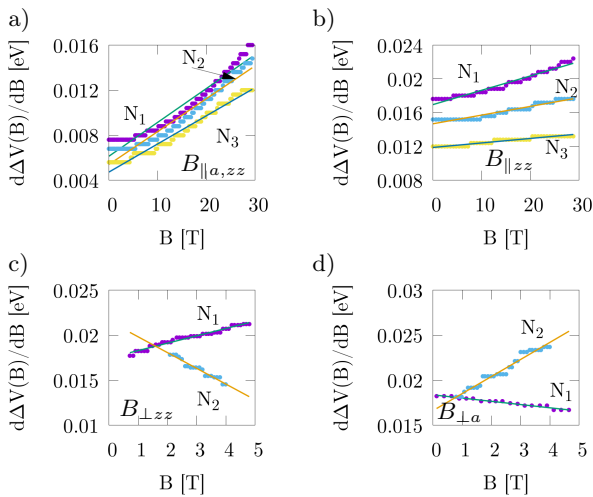
\begin{figure}[htbp]
\centering

\includegraphics[width=0.24\textwidth, clip, trim=0cm 0cm 0cm 0cm]{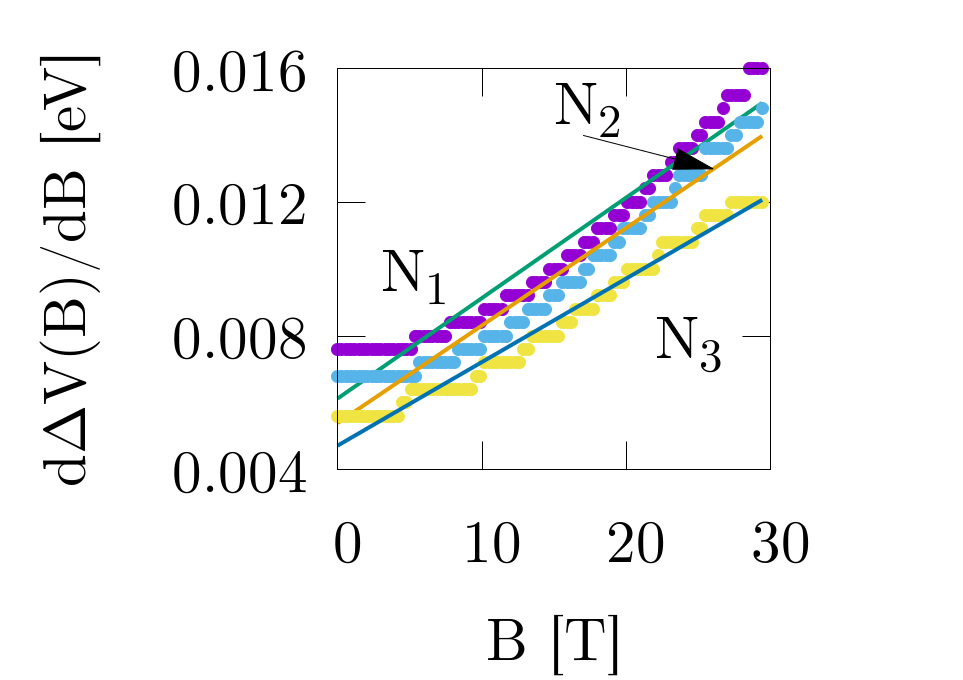}\includegraphics[width=0.24\textwidth, clip, trim=0cm 0cm 0cm 0cm]{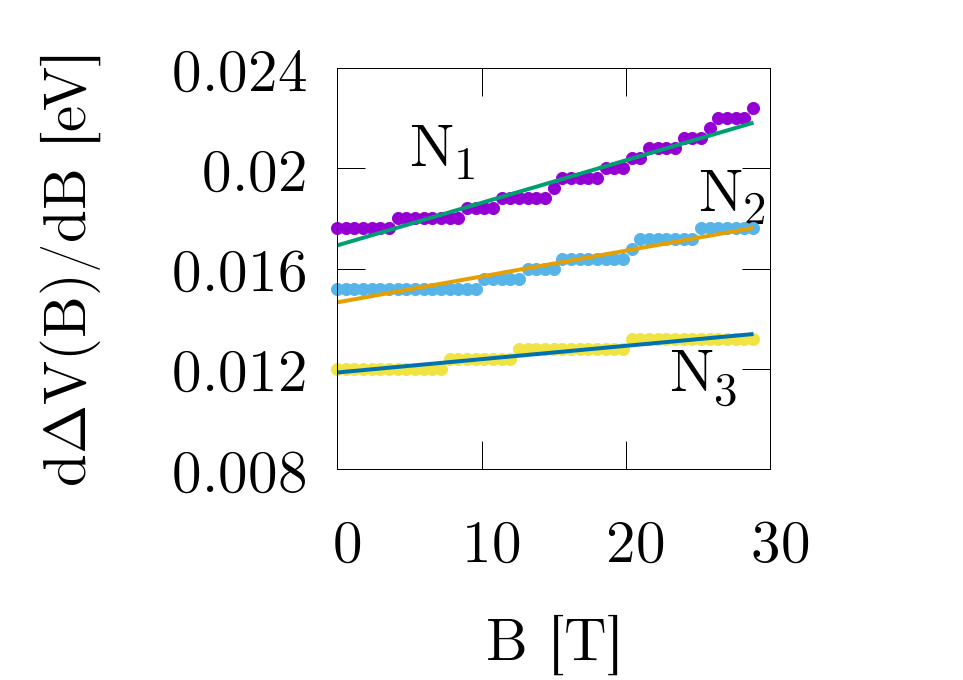}\put(-175,30){$B_{\| a,zz}$}\put(-60,32){$B_{\| zz}$}\put(-125,85){b)}\put(-240,85){a)}\\
\includegraphics[width=0.24\textwidth, clip, trim=0cm 0cm 0cm 0cm]{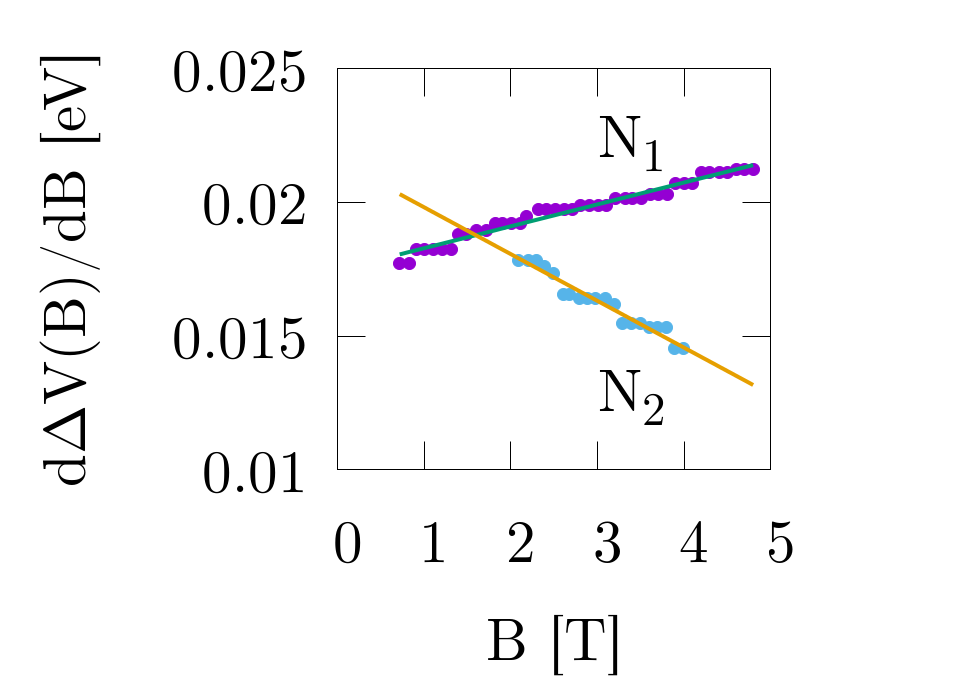}\includegraphics[width=0.24\textwidth, clip, trim=0cm 0cm 0cm 0cm]{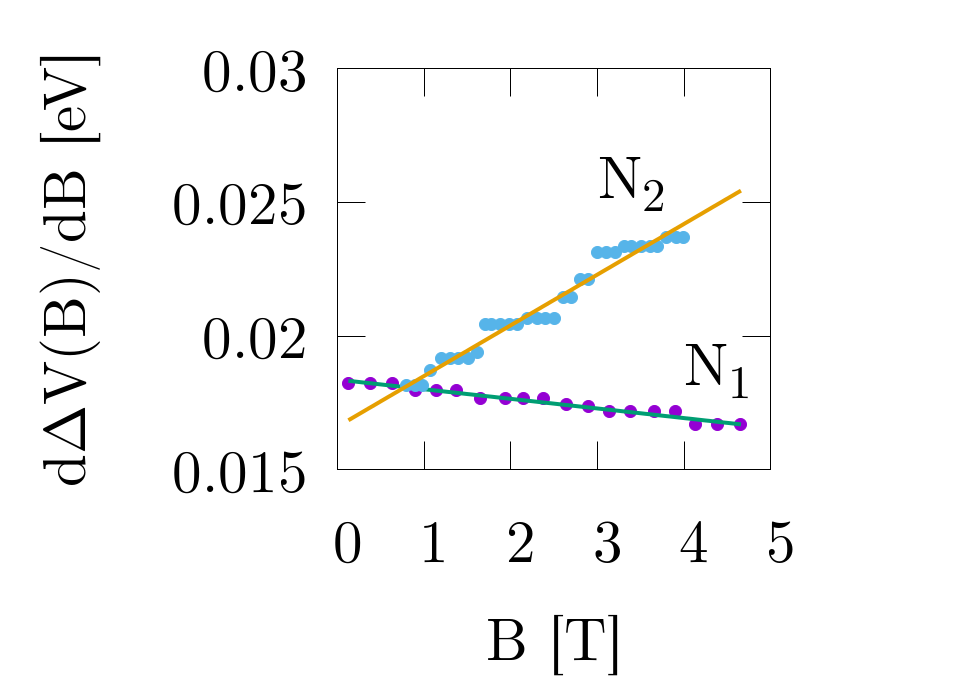}\put(-198,30){$B_{\bot zz}$}\put(-78,29){$B_{\bot a}$}\put(-125,85){d)}\put(-240,85){c)}
\caption{ The splitting of the transconductance for parallel magnetic field $B_\|$ (a) without SO and (b) with SO term active. For perpendicular $B_\bot$ (c) in zigzag and (d) in armchair nanoribbon. The slope of the fitted line for each split in the transconductance [see Fig.~\ref{fig:inplane_cond}(a,b)] is equal to $d(\Delta V_g(B))/dB$. }
\label{fig:fity}
\end{figure}

The second case concerns an in-plane magnetic field $B_\|$. We present the transconductance for the zigzag nanoribbon for $\mathbf{B} = [0,B_y,0]$ [Fig.~\ref{fig:inplane_cond}(a,b)] and we obtain similar results for $\mathbf{B} = [B_x,0,0]$ fields (not shown). For the armchair structure in $B_\|$ transconductance plots looks similar (not shown) and $g^*$ were calculated separately.  The {new states that enhance the conductance} at $E_F$ come in pairs of the same spin-type for $B_y>0$ [Fig.~\ref{fig:inplane_cond}(c-f)]. Again, the splitting at $B_y=0$ is an effect of SO in KM form, but contrary to Fig.~\ref{fig:zignoso}(a) this time we can calculate $g^*$ from transconductance even if double-states are visible -- the valley number in this case is not important. Slope of $\Delta V_g$ over $B_y$ is calculated from the fit (Fig.~\ref{fig:fity}) to the dashed lines in Fig.~\ref{fig:inplane_cond}(a,b). Results are presented in Tab.~\ref{tab:gstar}. Spin of an electron is strongly aligned along the $z$ axis when SO interaction is taken into account, hence the impact of external in-plane magnetic field is suppressed and we observe decreased $g^*<2$ values.

\begin{table}[h]
\caption{Effective Land\'{e} factors g*.}
\begin{tabular}{|c|c|c|c|}
\hline
SO off & $g^*_{1}$ & $g^*_{2}$ & $g^*_{3}$ \\ \hline
$B_\|$ &  2.1 & 2.2 & 2.1  \\ \hline
$B_\bot$ & 2.0 & 2.0 & 2.0  \\ \hline
\end{tabular}
\begin{tabular}{|c|c|c|c|}
\hline
SO on & $g^*_{1}$ & $g^*_{2}$ & $g^*_{3}$ \\ \hline
$B_{\| zz}$ & 1.2 & 0.78 & 0.45  \\ \hline
$B_{\| a}$ & 1.17 & 0.92 & 0.32  \\ \hline
$B_{\bot zz}$ & 5.8 & 13.3 & -  \\ \hline
$B_{\bot a}$ & 2.5 & 14.0 & -  \\ \hline

\end{tabular}
\label{tab:gstar}
\end{table}

\section{Summary and conclusions}

We studied the effective $g^*$ factors in electrostatic quantum point contacts defined in silicene using the tight-binding Hamiltonian by solving the scattering problem using the quantum transmitting boundary method. 
The spin-orbit coupling radically changes the values of the Land\'{e} factors. 
We showed that Zeeman splitting in magnetic field oriented parallel to the plane of the silicene lattice is isotropic and does not depend strongly on the edge type. Zeeman splitting from an external magnetic field is strongly suppressed by the intrinsic SO interaction in Kane-Mele form that introduces a Zeeman-like effective magnetic field perpendicular to the silicene plane. The spin-orbit interaction for the in-plane magnetic field decreases the effective $g^*$ factor to $g^*_1 = 1.2$ in the first subband, and  $g^*_2 = 0.78$, $g^*_3 = 0.48$ for the next two in the zigzag structure, respectively. In armchair nanoribbon we obtain similar results for $g^*$: $g^*_1 = 1.17$, moved slightly down/up/down (-0.02,+0.14,-0.13) compared to zigzag for the 3 first subbands, due to the mirrored $\nu_{kj}$ sign that adds a local magnetic field energy in KM form.

For the perpendicular orientation of the magnetic field we obtain effective Land\'{e} factor $g^*_1 = 5.8$ for the first subband and $g^*_2 = 13.3$ for the second in a zigzag nanoribbon, and $g^*_1 = 2.5$, $g^*_2 = 14.0$ for the armchair edge type. Reasoning remains the same as in parallel case but now the interaction of SO coupling is more visible in the $g^*$ factor for the first subbands, where local magnetic field significantly changes its value.


\section*{Acknowledgments}
B.R. is supported by Polish government budget for science in 2017-2021 as a research project under the program "Diamentowy Grant" (Grant No. 0045/DIA/2017/46), by the EU Project POWR.03.02.00-00-I004/16 and NCN grant UMO2019/32T/ST3/00044. A.M-K. is supported with "Diamentowy Grant" (Grant No. 0045/DIA/2017/46). The calculations were performed on PL-Grid Infrastructure
on Prometheus at ACK-AGH Cyfronet.

\bibliographystyle{apsrev4-1}
\bibliography{bib_silicene}

\end{document}